\documentclass[epj]{svjour}

\usepackage{graphicx}
\usepackage{fancyhdr}
\def\makeheadbox{{
 }}
\def\lsim{\mathrel{\raise.3ex\hbox{$<$\kern-.75em\lower1ex\hbox{$\sim$}}}}

\def\gsim{\mathrel{\raise.3ex\hbox{$>$\kern-.75em\lower1ex\hbox{$\sim$}}}}

\def\mytitle{My title} 
\def\myauthors{My name}  
\def\mytype{My type of session}
\def\mysession{My session}

\def\mytitle{Indirect Searches For Dark Matter: Signals, Hints and Otherwise} 
\def\myauthors{Dan Hooper}    
\def\mytype{Review}
\def\mysession{\myauthors}

\begin{document}
\title{Indirect Searches For Dark Matter}
\subtitle{(Signals, Hints and Otherwise)}
\author{Dan Hooper\inst{1}
}                     
\institute{Theoretical Astrophysics, Fermi National Accelerator Laboratory, Batavia, IL, USA}
%
\date{}
\abstract{For the SUSY 2007 conference, I was asked to review the topic of indirect searches for dark matter. As part of that talk, I summarized several observations which have been interpreted as the product of dark matter annihilations. In my contribution to the proceedings, I have decided to focus on this aspect of my talk. In particular, I will discuss the cosmic positron spectrum measured by HEAT, the 511 keV emission from the Galactic Bulge measured by INTEGRAL, the diffuse galactic and extragalactic gamma ray spectra measured by EGRET, and the microwave excess from the Galactic Center observed by WMAP. 
} 
\maketitle
\section{Introduction}
\label{intro}

The quest to uncover dark matter's particle identity is a multifaceted one. In addition to collider searches for stable, weakly interacting particles, a wide range of astrophysical searches for dark matter are underway. These astrophysical searches can be classified as direct detection experiments, which hope to observe the elastic scattering of dark matter particles with nuclei, and indirect experiments, which search for the annihilation products of WIMPs, including gamma rays, neutrinos, positrons, antiprotons, antideuterons and synchrotron radiation. 

Over the past several years, a number of observations have been interpreted as possible products of dark matter annihilations. Here, I take the opportunity to summarize and discuss five of these observations. After briefly reviewing the basic features of these signals, I will discuss the particle physics and astrophysics that is required to generate them through the process of dark matter annihilations.

\section{The HEAT Positron Excess}
\label{sec:2} 

WIMPs annihilating throughout the halo of the Milky Way can contribute electrons and positrons to the Galactic cosmic ray spectrum. Once injected into the local halo, electrons and positrons propagate under the influence of the Galactic magnetic field, gradually losing energy through synchrotron emission and through inverse Compton scattering with starlight and the cosmic microwave background. It has long been hoped that dark matter could be identified as an excess in the antimatter-to-matter ratio in the cosmic ray spectrum relative to that which is expected from astrophysical mechanisms.

\begin{figure}
\includegraphics[width=0.45\textwidth,height=0.45\textwidth,angle=0]{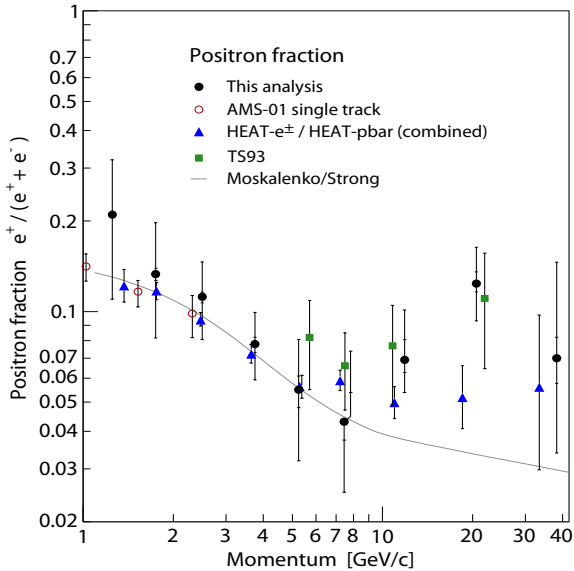}
\caption{The positron fraction measured by HEAT and AMS-01, as a function of energy. Above approximately 7 GeV, the measurements exceed the predictions of standard cosmic ray models~\cite{strongmoskalenko}. Figure taken from Ref.~\cite{ams01}. ``This analysis'' refers to the results of that paper.}
\label{fig:1}       
\end{figure}

In its balloon flights in 1994, 1995 and 2000, the HEAT experiment measured the positron and electron cosmic ray spectra at energies up to $\sim$30 GeV~\cite{heat}. When the measured position fraction ({\it ie.} the positron flux divided by the flux of electrons plus positions) is compared to the expectation from secondary production in cosmic ray propagation, an excess is seen at energies above approximately 7 GeV. A recent reanalysis of AMS-01 data appears to support this observation~\cite{ams01}. Neglecting uncertainties involved in the Galactic cosmic ray model, the combined statistical significance of this excess has been estimated at 4 to 5$\sigma$~\cite{45sigma}. In Fig.~\ref{fig:1} these measurements are shown and compared to the expected ratio from cosmic ray propagation alone.

Although the inclusion of a population of positrons produced through dark matter annihilations can improve the fits of the HEAT and AMS-01 data~\cite{positronheat} this is not particularly easy to accommodate within the context of the most simple models. For the case of a typical thermally produced WIMP ($\sigma v \lsim 3 \times 10^{-26}$ cm$^3$/s), the injected flux of positrons is too small by a factor of 50 or more to account for the observed excess~\cite{baltzpos}. If terms in the annihilation cross section proportional to $v^2$ or coannihilations play a significant role in the thermal freeze-out of the WIMP, then the annihilation rate in the local halo will be corresponding suppressed, further exacerbating this problem.  

Fluctuations in the distribution of dark matter can lead to enhancements in the average annihilation rate, known as the ``boost factor''. If this boost factor is sufficiently large, then a flux compatible with the HEAT signal could perhaps be generated. Assuming that there are no very large and unknown concentrations of dark matter within a few surrounding kiloparsecs (which, although not impossible, is very unlikely~\cite{hoopertaylorsilk}), however, it not generally expected that the boost factor would be large enough to account for the observed positron excess. That being said, such a scenario cannot be completely ruled out. Another possibility is that the dark matter consists of particles that were not produced thermally in the early universe, and therefore may have a considerably larger annihilation cross section, thus ameliorating the need for a large boost factor.

In any case, whether or not the positron flux observed by HEAT and AMS-01 is the result of annihilating dark matter will become much more clear in the near future, as the positron and electron cosmic ray spectra are measured by the PAMELA satellite~\cite{pamela} which began its mission last year~\cite{silkpos}. This experiment will study the cosmic positron spectrum with much greater precision and to much higher energies than either HEAT or AMS-01. The dark matter community eagerly awaits their first results.

\section{The INTEGRAL 511 keV Line}
\label{sec:3} 

Four years ago, the SPI spectrometer on board the INTEGRAL satellite confirmed the very bright emission of 511 keV photons from the region of the Galactic Bulge, corresponding to an injection rate of approximately $3 \times 10^{42}$ positrons per second in the inner Galaxy~\cite{integral}. This is orders of magnitude larger than the expected rate from pair creation via cosmic ray interactions with the interstellar medium. The signal appears to be approximately spherically symmetric (with a full-width-half-maximum of approximately 6$^{\circ}$), with little of the emission tracing the Galactic Disk. Any stellar origin of this signal, such as type Ia supernovae, hypernovae or gamma ray bursts, would therefore also require a mechanism (such as substantial coherent magnetic fields) by which the positrons can be transported from the disk to throughout the volume of the Bulge~\cite{Prantzos:2005pz}. Furthermore, type Ia supernovae do not inject enough positrons to generate the observed intensity of this signal~\cite{Kalemci:2006bz}. It is possible, however, that hypernovae~\cite{Casse:2003fh} or gamma ray bursts~\cite{Casse:2003fh,Parizot:2004ph} might be capable of injecting positrons at a sufficient rate. Overall, it is difficult to explain the intensity and morphology of the 511 keV emission with astrophysical mechanisms.

In light of the challenges involved in explaining the 511 keV emission from the Bulge, Celine Boehm and collaborators (including myself) suggested that this signal could potentially be the product of dark matter annihilations~\cite{511dark}. In order for dark matter particles to generate the observed spectral line width of this signal, however, their annihilations must inject positrons with energies below a few MeV~\cite{beacom}. This, in turn, implies that the dark matter's mass be near the 1-3 MeV range -- much lighter than annihilating dark matter particles in most models.

Although weakly interacting particles with masses smaller than a few GeV (but larger than $\sim$1 MeV) tend to be overproduced in the early universe relative to the measured dark matter abundance~\cite{lee}, this can be avoided if a new light mediator is introduced which makes dark matter annihilations more efficient~\cite{lightok}. For example, although neutralinos within the MSSM are required by relic abundance considerations to be heavier than $\sim$20 GeV~\cite{susycase}, they can be much lighter in extended supersymmetric models in which light Higgs bosons can mediate neutralino annihilations~\cite{nmssm}.

For dark matter particles with MeV-scale masses to generate the measured dark matter abundance, they must annihilate during the freeze-out epoch with a cross section of $\sigma v \sim 3 \times 10^{-26}$ cm$^3$/s. To inject the flux of positrons needed to generate the signal observed by SPI/INTEGRAL, however, an annihilation cross section four to five orders of magnitude smaller is required. Together, these requirements force us to consider dark matter particles with s-wave suppressed annihilations (dominated by p-wave contributions, $\sigma v \propto v^2$). Such behavior can be found, for example, in the case of fermionic or scalar dark matter particles annihilating through a vector mediator. For such a dark matter particle to not be overabundant today, the mediating boson also must be quite light~\cite{scalar}.

As an alternative explanation for the 511 keV line, it has recently been proposed that $\sim 500$ GeV dark matter particles could be collisionally excited to a 1-2 MeV heavier state, followed by the de-excitation to the ground state plus an electron-positron pair, thus generating the positrons needed to account for the observed 511 keV signal~\cite{exciting}.

\section{EGRET's Diffuse Galactic Spectrum}
\label{sec:4}

\begin{figure*}
\includegraphics[width=0.33\textwidth,height=0.34\textwidth,angle=0]{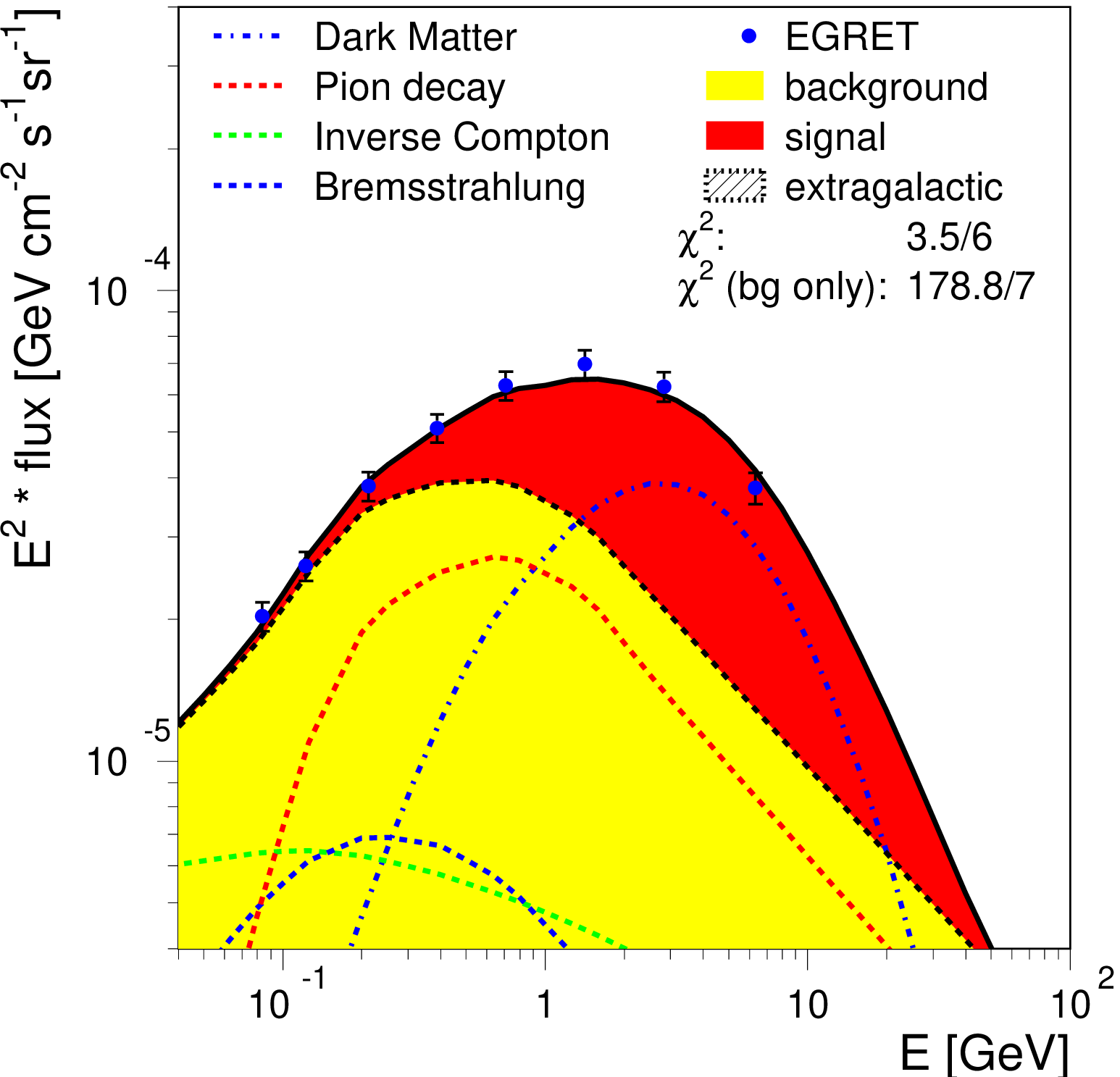}
\includegraphics[width=0.33\textwidth,height=0.34\textwidth,angle=0]{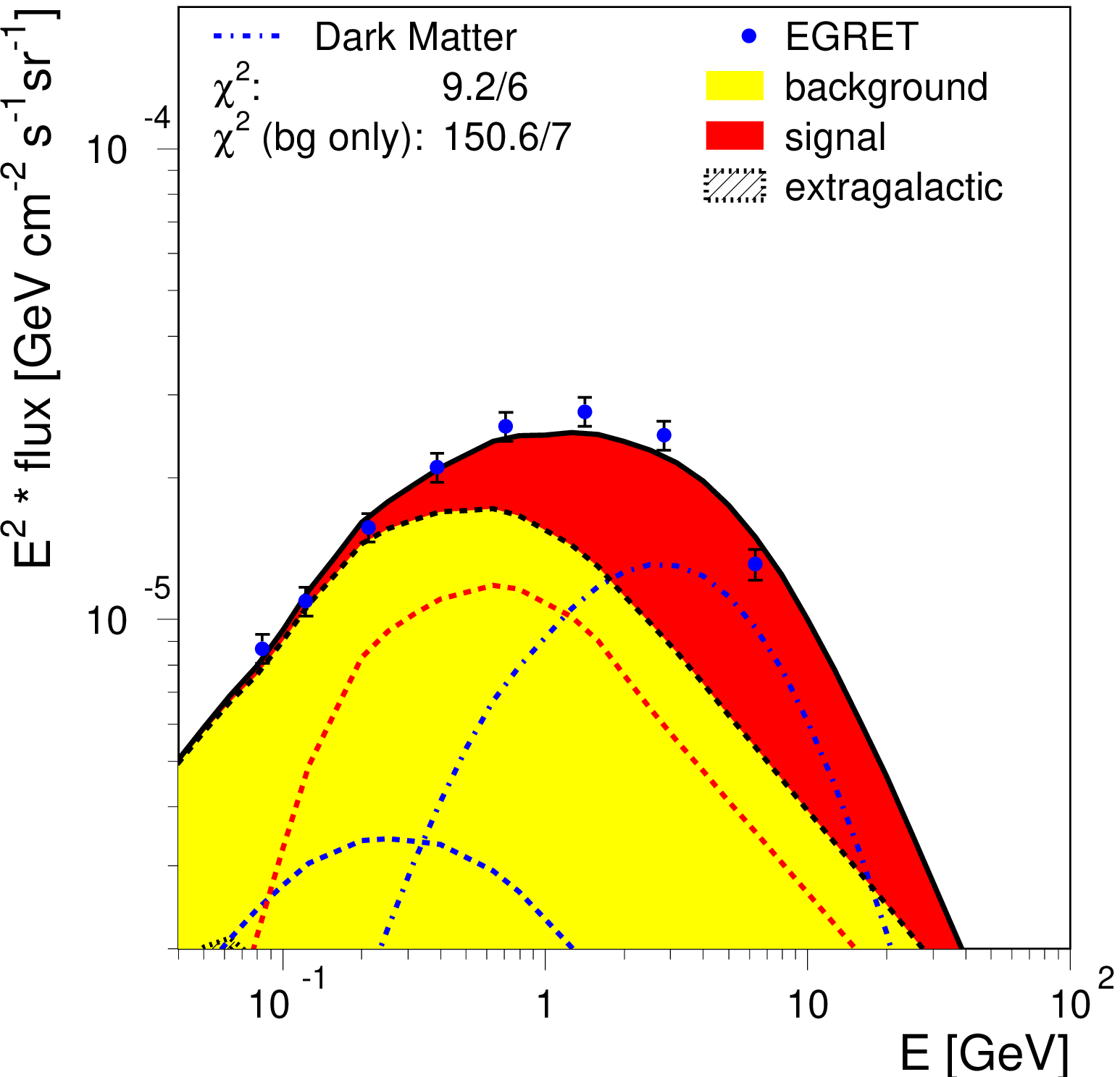}
\includegraphics[width=0.33\textwidth,height=0.34\textwidth,angle=0]{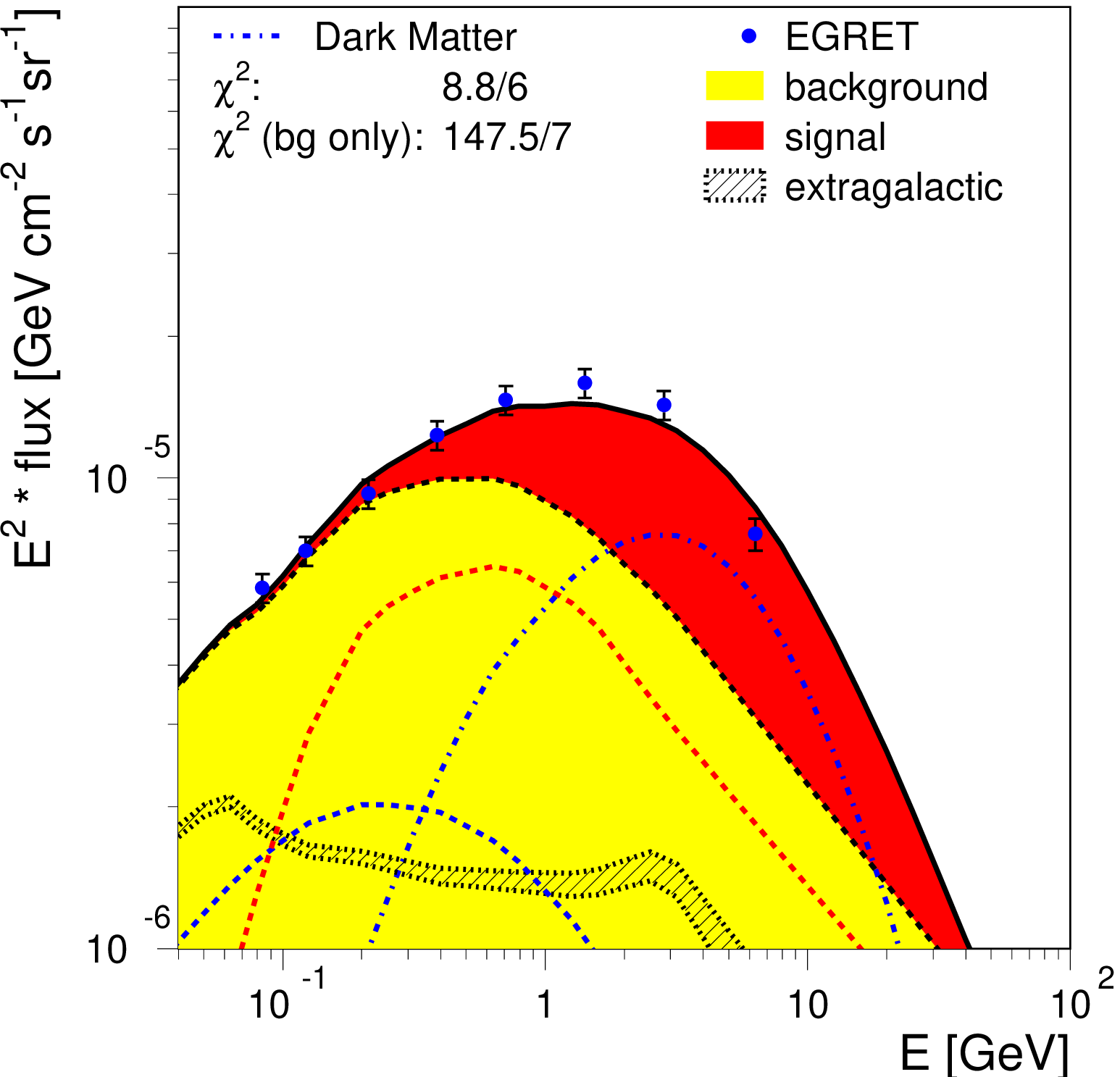}
\\
\includegraphics[width=0.33\textwidth,height=0.34\textwidth,angle=0]{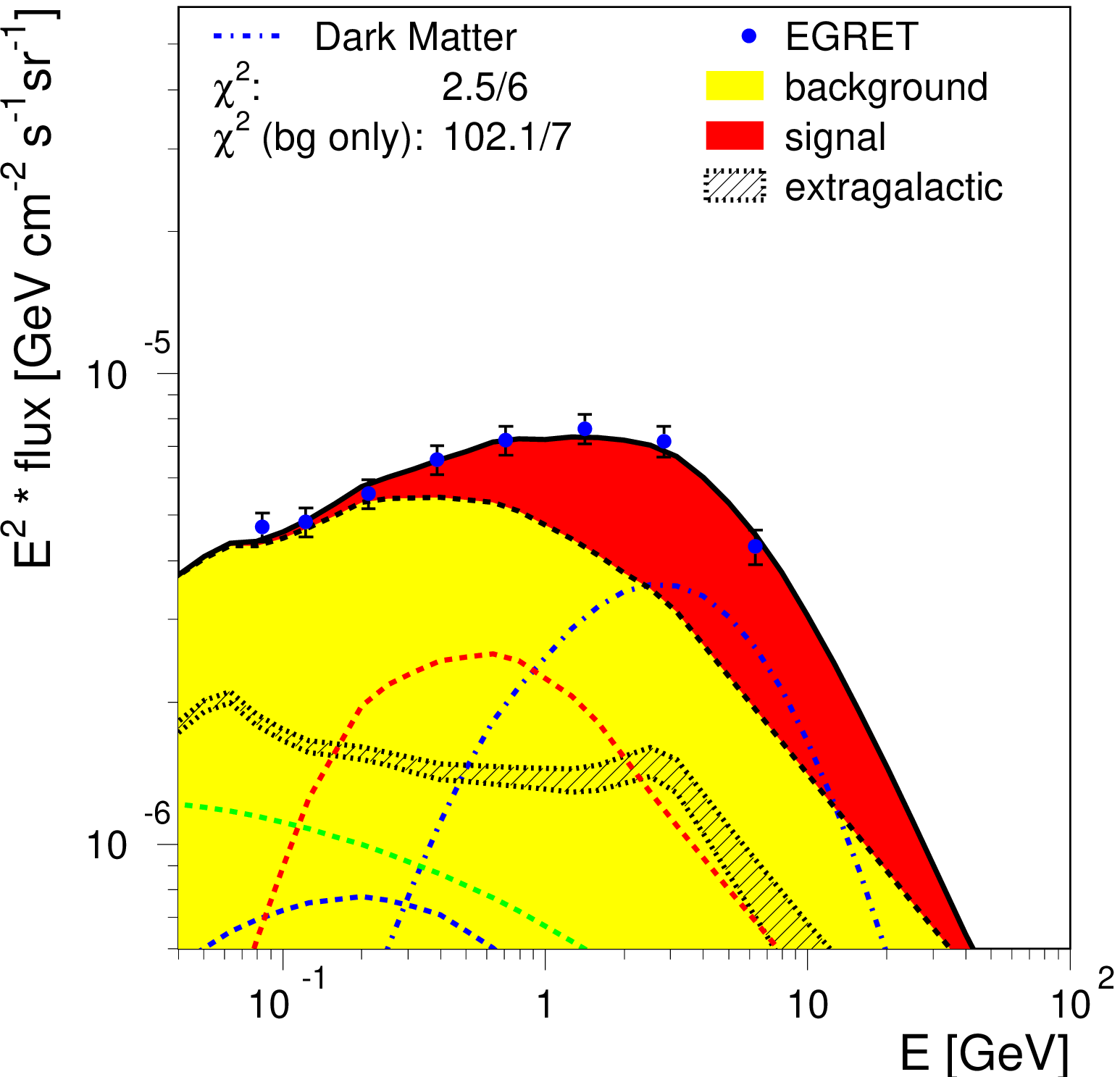}
\includegraphics[width=0.33\textwidth,height=0.34\textwidth,angle=0]{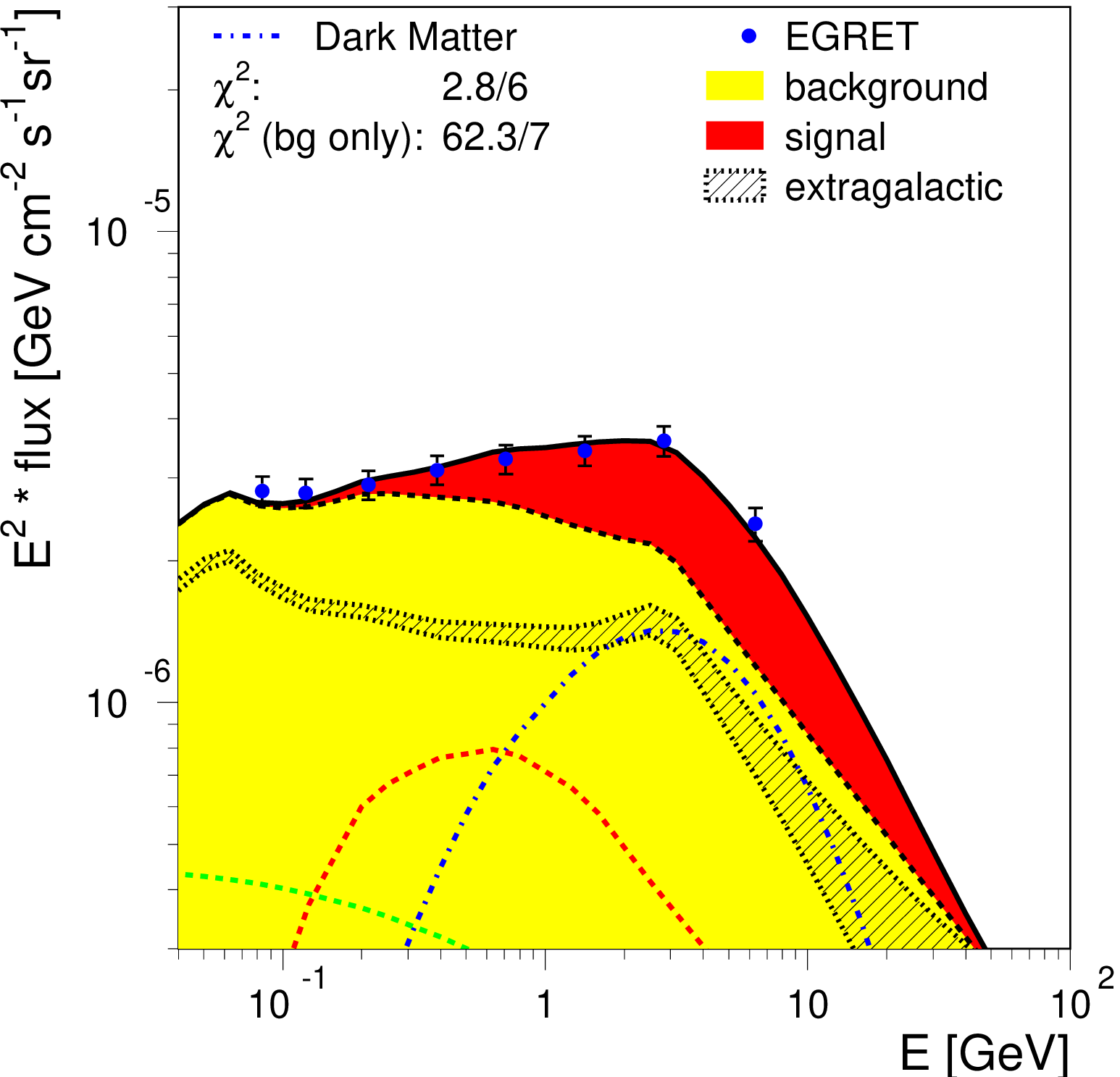}
\includegraphics[width=0.33\textwidth,height=0.34\textwidth,angle=0]{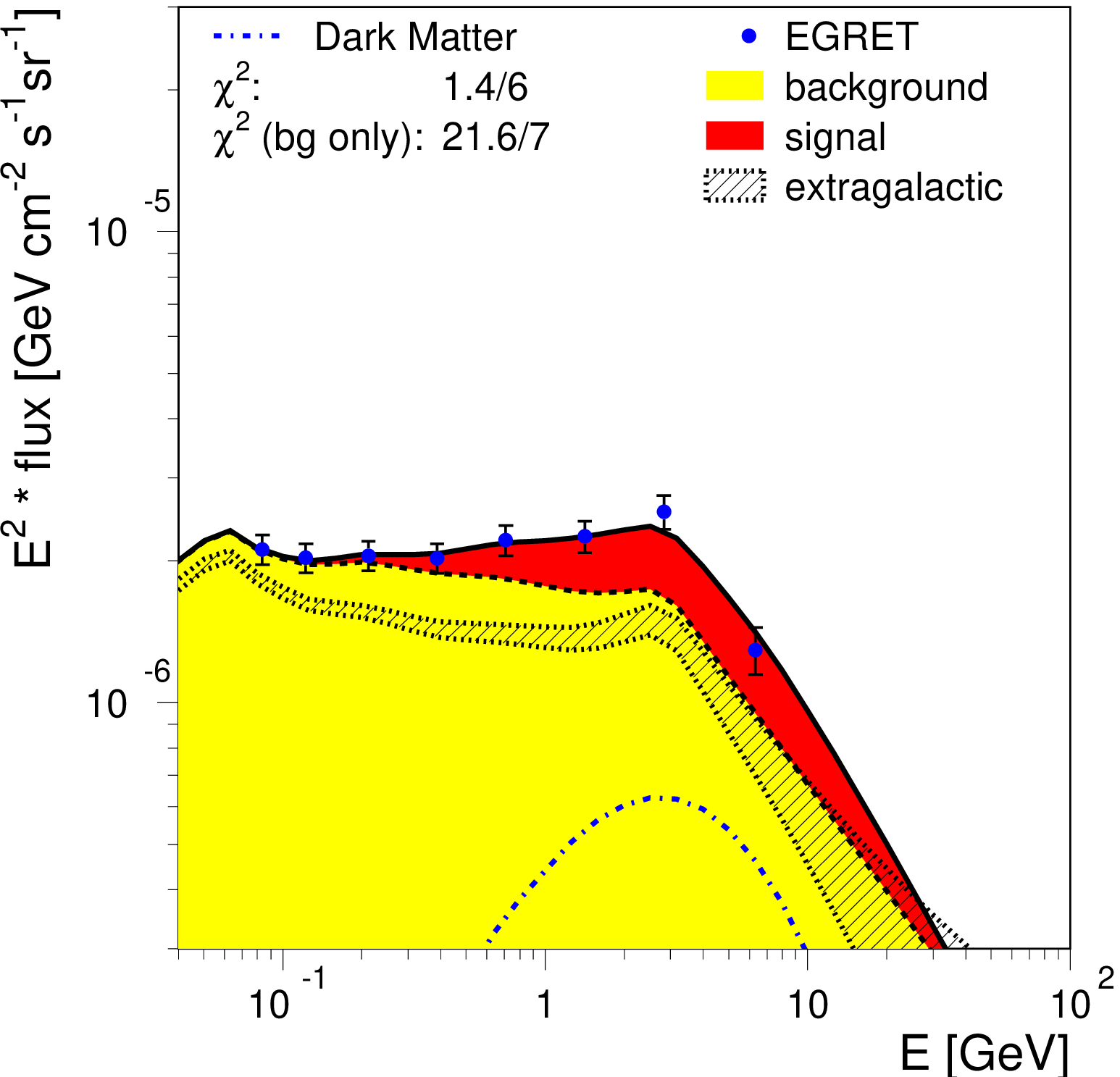}
\caption{Fits to the diffuse gamma ray spectrum as measured by EGRET. The six frames represent six different regions of the sky. The inclusion of gamma rays from the annihilations of a 60 GeV WIMP (red) improve the fit considerably in each of the regions. From Ref.~\cite{deboer}}
\label{fig:2}       
\end{figure*}

The satellite-based gamma ray detector, EGRET, has measured the diffuse spectrum of gamma rays over the entire sky. When compared to conventional Galactic models, these measurements appear to contain an excess at energies above approximately 1 GeV. This has been interpreted as evidence for dark matter annihilations in the halo of the Milky Way~\cite{deboer}. 

Among the most intriguing features of the observed EGRET excess is its similar spectral shape over all regions of the sky. Furthermore, this shape is consistent with that predicted from annihilations of a 50-100 GeV WIMP. In Fig.~\ref{fig:2} the background and dark matter annihilation spectra are shown and compared to the EGRET measurements over various regions of the sky.

There are, however, some substantial challenges involved with interpreting the EGRET excess as a product of dark matter annihilations. In particular, the normalization of the dark matter annihilation rate in each of the six frame of Fig.~\ref{fig:2} was selected independently. To accommodate the required normalization throughout the Galaxy, the distribution of dark matter has to depart substantially from the predictions of standard dark matter halo profiles. In particular, Ref.~\cite{deboer} adopts a distribution which includes two very massive ($\sim 10^{10}\, M_{\odot}$) toroidal rings of dark matter near or within the Galactic Plane, at distances of approximately 4 and 14 kiloparsecs from the Galactic Center. The authors motivate the presence of these rings by observed features in the Galactic rotation curve, and suggest that they may be remnants of very massive dwarf galaxies which have been tidally disrupted.

\begin{figure}
\includegraphics[width=0.45\textwidth,height=0.45\textwidth,angle=0]{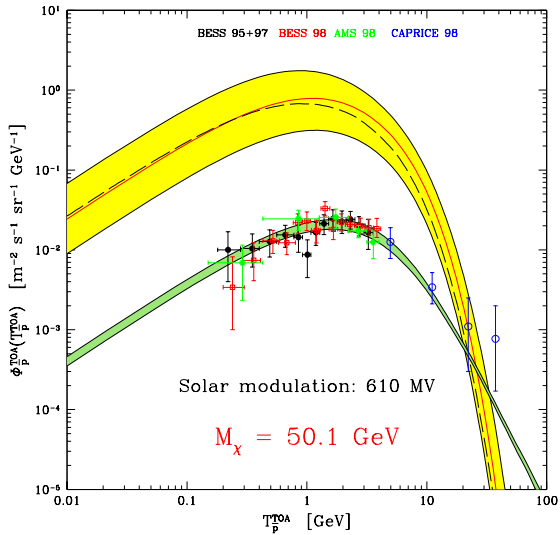}
\caption{If the diffuse Galactic gamma ray excess observed by EGRET is produced through dark matter annihilations, then a large flux of cosmic antiprotons is also expected. The spectrum of cosmic antiprotons predicted in such a scenario, assuming standard cosmic ray diffuse models, is shown as the yellow band. This range of predictions is clearly in disagreement with the observed antiproton spectrum. In contrast, the predictions of standard cosmic ray models (without dark matter), which is shown as the green band, match the data quite well. From Ref.~\cite{bergstromdeboer}.}
\label{fig:3}       
\end{figure}

The other difficulty involved with the interpretation of the EGRET excess as dark matter annihilation radiation is the large flux of antiprotons which is expected to be generated in such a scenario~\cite{bergstromdeboer}. In particular, the flux of cosmic antiprotons produced is expected to exceed the measured flux by more than an order of magnitude (see Fig.~\ref{fig:3}). To avoid this conclusion, one is forced to consider significant departures from standard galactic diffusion models. In particular, an anisotropic diffusion model featuring strong convection away from the Galactic Disk and a large degree of inhomogeneities in the local environment could reduce the cosmic antiproton flux to acceptable levels~\cite{deBoer:2006ck}.

The dark matter interpretation of EGRET's measurment of the galactic diffuse spectrum has also been challenged on the grounds that the data could plausibly be explained without the addition of an exotic component, from dark matter or otherwise. In particular, uncertainties in the cosmic ray propagation and diffusion model lead to considerable variations in the predicted diffuse gamma ray backgrounds~\cite{Moskalenko:2006zy}. More recently, it has also been suggested that the observed excess could also be the result of systematic errors in EGRET's calibration~\cite{Stecker:2007xp}.

\section{EGRET's Diffuse Extragalactic Spectrum}
\label{sec:5}

In this section, I will discuss another aspect of the EGRET data which has been interpreted as a possible product of dark matter annihilations. In particular, is has been proposed that dark matter annihilation radiation may constitute a significant fraction of the extragalactic (isotropic) diffuse gamma ray flux~\cite{Elsaesser:2004ap}.

\begin{figure}
\includegraphics[width=0.45\textwidth,height=0.45\textwidth,angle=0]{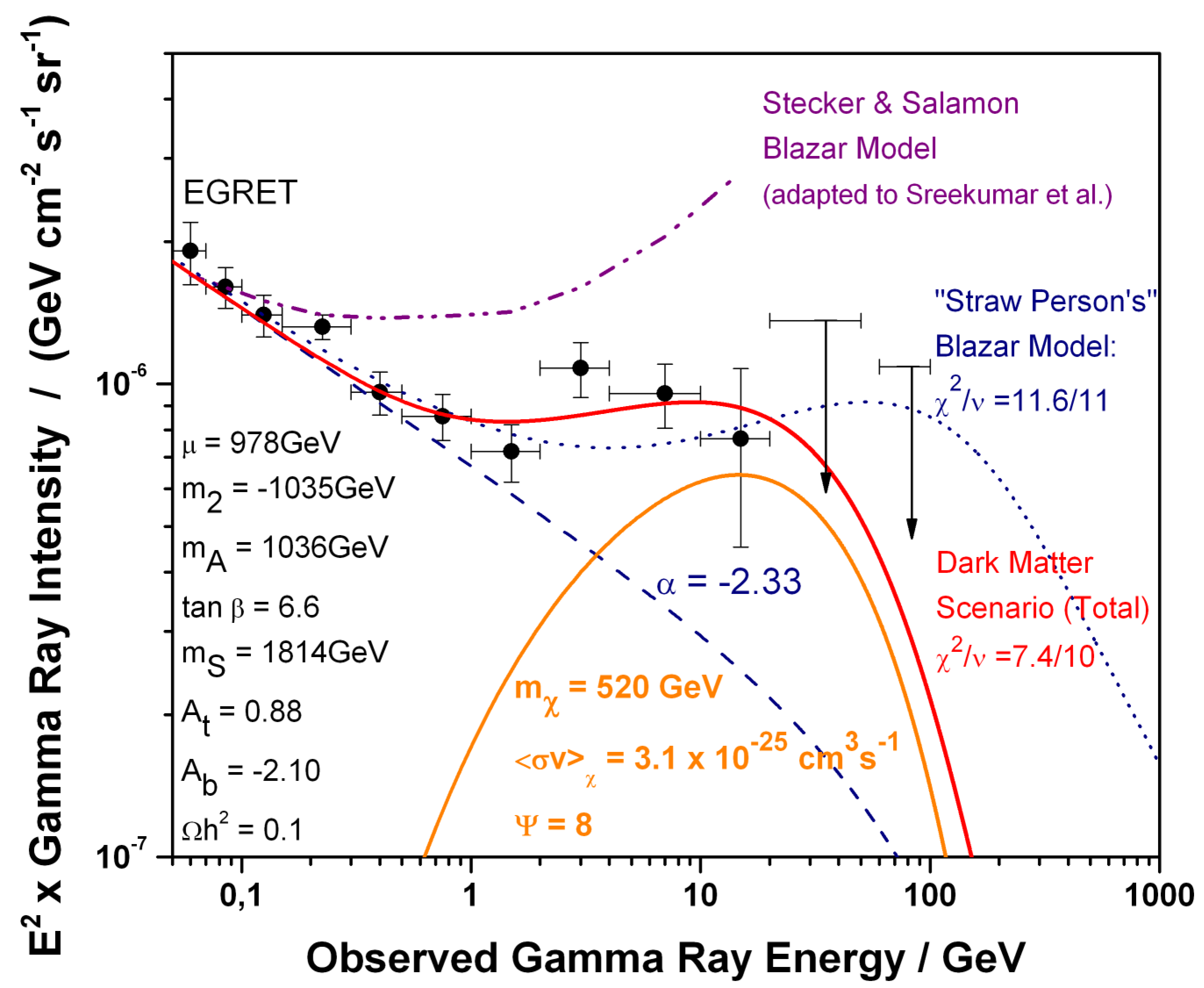}
\caption{The isotropic, extragalactic gamma ray background as observed by EGRET compared to the predictions of various models. A contribution from a roughly 500 GeV WIMP fits the data reasonably well. From Ref.~\cite{Elsaesser:2004ap}.}
\label{extrag}       
\end{figure}

In Fig.~\ref{extrag}, the diffuse extragalactic gamma ray background, as measured by EGRET, is shown and compared to the predictions of various models. Although all or most of the observed spectrum could be the product of astrophysical source such as blazars, much of the flux observed in the 1-20 GeV range could also be the result of dark matter annihilations taking place throughout the universe~\cite{ullio}. In particular, the observed spectrum fits reasonably well the predictions for a WIMP with a mass of roughly 500 GeV.

The dark matter annihilation rate needed to normalize to the diffuse flux measured by EGRET is, however, quite high and requires either a very large dark matter annihilation cross section or dark matter halos which are very cuspy. In particular, if an Navarro-Frenk-White profile~\cite{nfw} is adopted for all halos throughout the universe, then an annihilation cross section $10^2$ to $10^3$ times larger than is predicted for a thermal relic is required to generate the observed gamma ray flux. If the dark matter distribution in the Milky Way is similar to that found in halos throughout the universe, however, then the gamma ray flux from the center of our galaxy would far exceed that which is observed~\cite{Ando:2005hr}. To generate the isotropic diffuse flux observed by EGRET without conflicting with observations of the Galactic Center, therefore, requires extremely cusped halo profiles in most or at least many of the galaxies throughout the universe, and a far less dense cusp in our own Milky Way.

\section{The WMAP Haze}
\label{sec:6} 

In addition to its measurements of the cosmic microwave background, data from the Wilkinson Microwave Anisotropy Probe (WMAP) has been used to provide the best measurements to date of the standard interstellar medium emission mechanisms, including thermal dust, spinning dust, ionized gas, and synchrotron. In addition to these expected foregrounds, the observations have revealed an excess of microwave emission in the inner $20^{\circ}$ around the center of the Milky Way, distributed with approximate radial symmetry.  This excess is known as the ``WMAP Haze''~\cite{haze1}.

Although the WMAP Haze was initially thought likely to be thermal bremsstrahlung (free-free emission) from hot gas ($10^4 \, \rm{K}\gg T \gg 10^6\, \rm{K}$), this interpretation has since been ruled out by the absence of an H$\alpha$ recombination line and X-ray emission. Other possible origins for this signal, such as thermal dust, spinning dust, and Galactic synchrotron as traced by low-frequency surveys, also seem unlikely. More recently, it has been suggested that
the WMAP Haze, could be generated as a product of dark matter annihilations~\cite{hazedark1}. In particular, annihilating dark matter particles produce relativistic electrons and positrons which travel under the influence of the Galactic magnetic field. As they do, they will emit synchrotron photons, which naturally fall within the frequency range measured by WMAP.

\begin{figure}
\includegraphics[width=0.47\textwidth,height=0.33\textwidth,angle=0]{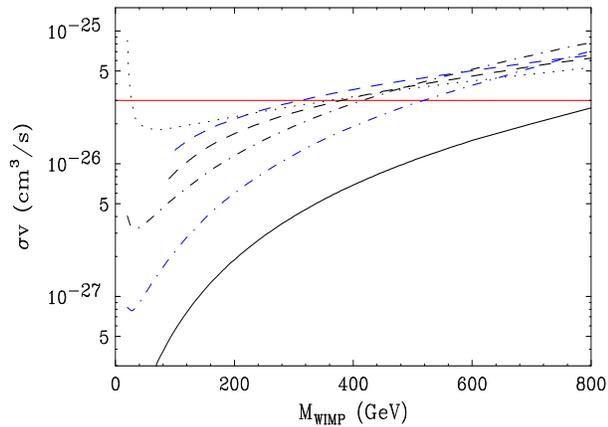}
\caption{The WIMP annihilation cross section required to produce the observed intensity of the WMAP Haze, as a function of the WIMP mass. Each contour denotes a different dominant annihilation mode. From top-to-bottom (on the left): $b\bar{b}$, $ZZ$, $W^+W^-$, $\tau^+\tau^-$, $\mu^+\mu^-$, $e^+ e^-$. Note that cross sections near the value predicted of a typical thermal relic ($3\times 10^{-26}$cm$^3$/s) are found. No boost factors or other enhancements to the annihilation rate are required to generate the observed signal.}
\label{sigma}       
\end{figure}

The angular distribution of the Haze can be used to constrain the shape of the required dark matter halo profile. In Ref.~\cite{hazedark2}, it was shown that the Haze is consistent with dark matter distributed as $\rho(r) \propto r^{-1.2}$ within the inner several kiloparsecs of our galaxy. This slope falls between those predicted by the Navarro-Frenk-White, $\rho(r) \propto r^{-1}$, and Moore {\it et al.}, $\rho(r) \propto r^{-1.5}$, halo profiles. Once the dark matter's halo profile has been fixed by the angular distribution of the Haze, the intensity of the signal can be used to determine the required annihilation cross section. In Fig.~\ref{sigma}, the WIMP annihilation cross section required to produced the observed intensity of the WMAP Haze is shown for several possible annihilation modes, as a function of the WIMP mass. Intriguingly, a 50-1000 GeV WIMP annihilating to heavy fermions or gauge bosons generates synchrotron emission with an intensity within a factor of two or three of that which is observed in the WMAP Haze. No boost factors or exotic astrophysical parameters are required to generate this signal through dark matter annihilations.

It is also interesting to note that the dark matter halo profile and annihilation cross section required to generate the WMAP Haze through dark matter annihilations imply a flux of prompt gamma rays from the Galactic Center region that is within the reach of the upcoming GLAST experiment~\cite{Hooper:2007gi}. Additionally, the upcoming Planck satellite will provide substantially improved measurements of the spectrum of the Haze.

\section{Comparisons, Discussion and Summary}
\label{summary}

I have summarized here five different astrophysical signals which have been interpreted as possible products of dark matter annihilations. Each of these has their own strengths and weaknesses. Attempting to remain as objective as possible, I would like offer my own assessment of these various observations.

\begin{table*}
\caption{A comparison of the particle physics and astrophysics required in each of the five scenarios I have discussed.}
\label{tab:1}       
\begin{tabular}{lll}
\hline\noalign{\smallskip}
Signal & Required Particle Physics & Required Astrophysics  \\
\noalign{\smallskip}\hline\noalign{\smallskip}
HEAT positron spectrum &Non-thermal production mechanism or... & Boost factor of 50 or more  \\
& Electroweak-scale WIMP & \\
\hline\noalign{\smallskip}
INTEGRAL 511 keV line & MeV mass dark matter particle & Mildly cusped halo profile \\
& p-wave annihilation cross section & \\
\hline\noalign{\smallskip}
Diffuse galactic $\gamma$-rays & $\sim$50-100 GeV WIMP & Two massive dark matter rings \\
&&Anisotropic diffusion (convention, etc.) \\
\hline\noalign{\smallskip}
Diffuse extragalactic $\gamma$-rays & $\sim$500 GeV WIMP & Dense cusps common  \\
&& No cusp in the Milky Way  \\
\hline\noalign{\smallskip}
WMAP Haze & Electroweak-scale WIMP & Cusped halo profile  \\
&Cross section predicted for a thermal relic & No boost factors/exotic astrophysics  \\
\noalign{\smallskip}\hline
\end{tabular}
\vspace*{1cm}  
\end{table*}

In Table~\ref{tab:1}, I have summarized the nature of the particle physics and astrophysics which must be introduced to generate the observed signals in each of the five scenarios I have discussed. Based on these requirements, I would like to make the following points:

\begin{itemize}

\item{In the case of the positron excess and each of the two EGRET excesses, there is a significant degree of difficulty involved in generating an annihilation rate sufficiently large to produce the observed signal. As a result, some combination of large annihilation cross sections or cusped profiles (or boost factors) are required. That being said, such rates might be possible, even though they are higher than our naive expectations suggest.}

\item{The two EGRET signals each require somewhat more complicated astrophysical setups if they are to be interpreted as products of dark matter annihilation. In the case of the Galactic excess, the diffusion model has to be modified to avoid the overproduction of antiprotons. Also, a rather exotic (although not impossible) dark matter distribution also has to be adopted. Regarding the extragalactic signal, the annihilation rate has to be very large in most galaxies, but not in the Milky Way. While not impossible, it requires us to find ourselves living somewhere marginally exceptional.}

\item{The INTEGRAL signal stands out among these five in that it requires a very different dark matter particle than is generally considered. The conventional wisdom is that it is not easy to construct a viable particle physics model containing an MeV dark matter candidate. I leave it to you to judge the merits of this argument.}

\item{The WMAP Haze stands alone among these five signals in that it requires neither unexpected particle physics or astrophysics in order to be generated through dark matter annihilations. The observed angular distribution, intensity and spectrum are consistent with a vanilla electroweak-scale particle which was produced thermally in the early universe and is distributed with a cusped halo profile in our Galaxy.}

\item{It is worth noting that {\it none} of the signals I have discussed here are particularly distinctive -- there is not yet a ``smoking gun''. Further observations will be required to draw any conclusive statements. Fortunately, new data from PAMELA, GLAST, Planck and even the LHC are not too far off. In a few years time, most (if not all) of these signals will have either gone away or become much more interesting.}

\end{itemize}

%
%

\end{document}